\documentclass[a4paper]{PoS}

\title{Electroweak production of Higgs boson pairs in 2HDMs}

\ShortTitle{Electroweak production of Higgs boson pairs in 2HDMs}

\author{Rikard Enberg\\
        Department of Physics and Astronomy, Uppsala University, 
Box 516, SE-751 20 Uppsala, Sweden\\
        E-mail: \email{rikard.enberg@physics.uu.se}}

\author{William Klemm\\
        School of Physics \& Astronomy, University of Manchester, 
Manchester M13 9PL, UK \\
        E-mail: \email{william.klemm@physics.uu.se}}

\author{Stefano Moretti\\
       School of Physics \& Astronomy,
University of Southampton, Southampton SO17 1BJ, UK \\
       E-mail: \email{s.moretti@soton.ac.uk}}

\author{\speaker{Shoaib Munir}\\
        School of Physics, Korea Institute for Advanced Study,
Seoul 130-722, Republic of Korea \\
        E-mail: \email{smunir@kias.re.kr}}

\abstract{One of the main features of a Two-Higgs Doublet Model (2HDM)
  is the presence of two additional neutral Higgs states, besides the
  one mimicking the $\sim125$\,GeV state observed at the LHC. The
  three Higgs bosons of a 2HDM can be produced at the LHC either singly via
  gluon fusion or in pairs with each other. When analyzing their pair
  production, the emphasis is laid on
  gluon-initiated processes, and the electroweak (EW) production is generally
  not treated on the same footing, assuming its contribution to be highly
  subleading. We show here that when the sum of the masses of the lightest
  scalar and pseudoscalar Higgs bosons in the Type-I 2HDM is smaller
  than the $Z$-boson mass, their EW pair production can dominate over
  QCD pair production by orders of magnitude.}

\FullConference{38th International Conference on High Energy Physics\\
		3-10 August 2016\\
		Chicago, USA}

\begin{document}

\section{Light Higgs bosons in the Type-I 2HDM}

In the Type-I 2HDM only one of the two Higgs doublets, $\phi_1$ and $\phi_2$,
couples to all the Standard Model (SM) fermions, with a $Z_2$ symmetry
preventing large flavor changing neutral currents. The model contains
three neutral Higgs states, two scalars, $h$ and $H$, with $m_h < m_H$,
and a pseudoscalar, $A$. Either one of $h$ or $H$ can
play the role of the SM-like Higgs boson, $h_{\rm obs}$, discovered at
the LHC\ \cite{Aad:2012tfa, Chatrchyan:2012xdj}. In the scenario when
the mass and signal
rates of $H$ are consistent with those of $h_{\rm obs}$, $h$ can be as light as a
few GeV, without violating the
constraints from negative searches at the LEP
collider, Tevatron and LHC. When the $A$ is additionally light enough
that $m_h+m_A<m_Z$, their pair-production
via a resonant $Z$ in the $s$-channel becomes possible, but only in
the $q\bar{q}$-fusion process, since it is prohibited in the
gluon-fusion 
process by the Landau-Yang theorem\ \cite{Landau:1948kw,Yang:1950rg}. 
As a result, the production cross section of the $hA$ pair gets considerably
enhanced below the $Z$ mass.

\section{Numerical analysis}

\begin{figure}[b!]
\begin{center}
    \includegraphics[scale=0.55]{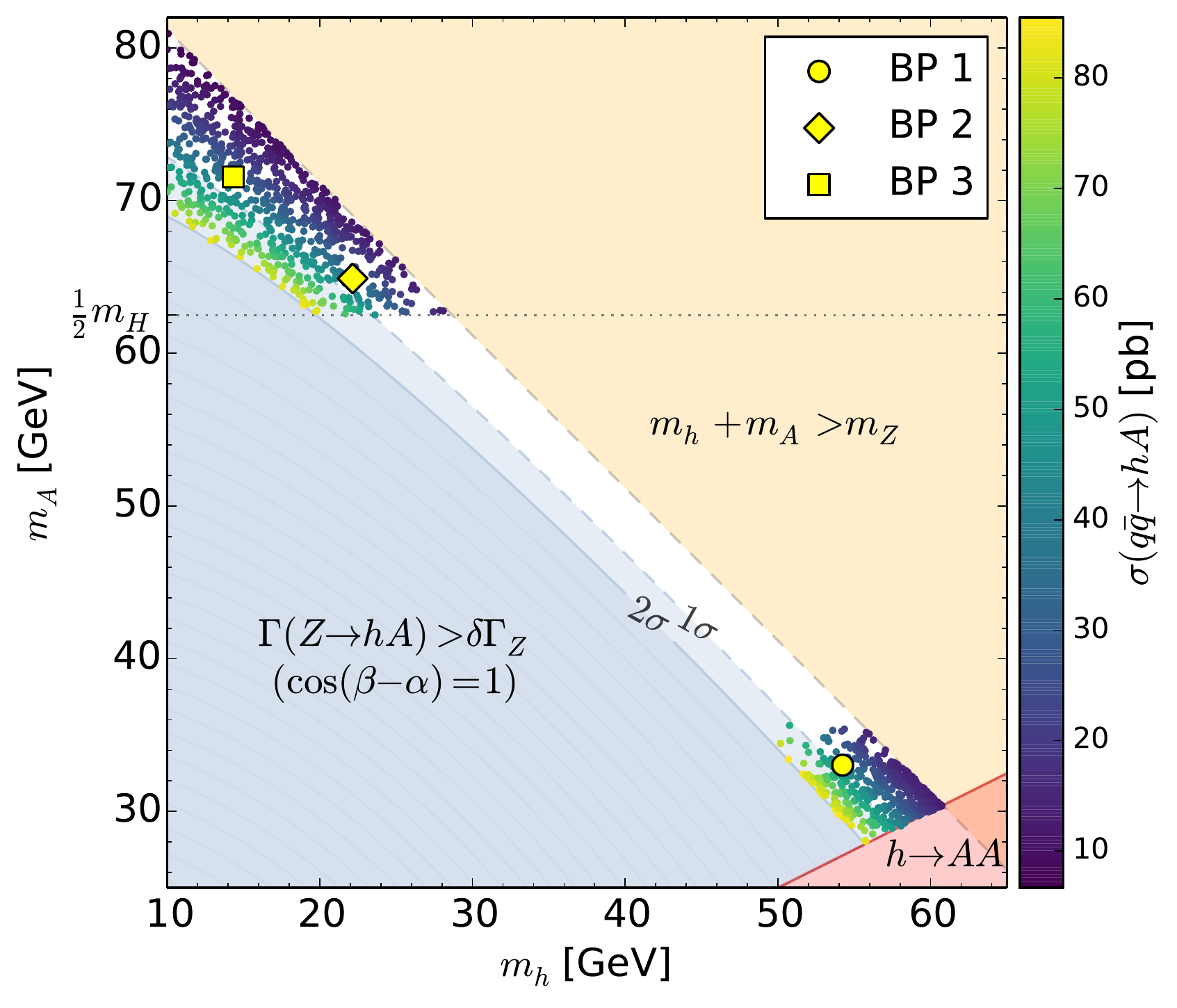}
    \caption{Points satisfying all the constraints imposed during the
      scan and additionally lying within
the experimental uncertainty on the $Z \to h A$ partial width, at the 
$1\sigma$ (lighter) and $2\sigma$
(darker) levels, assuming $\cos(\beta-\alpha)=1$.  
The small red region corresponds to $m_h > 2m_A$, allowing $h\to A A$
decays. The three benchmark points 
have been highlighted in yellow, and the
color map corresponds to the total cross section.
for the $q\bar{q}\to hA$ process at $\sqrt{s}=13$\ TeV.}
    \label{fig:ZWidth}
\end{center}
\end{figure}

To analyse the significance of the EW $hA$ pair-production, we first
performed a numerical scan of the six free parameters of the Type-I
2HDM using the 2HDMC-v1.7.0\ \cite{Eriksson:2009ws} program, in order to find points with $m_h+m_A<m_Z$ that are consistent 
with the results from collider searches as well as from $b-$physics and
EW precision experiments. These parameters include
$m_h,\,m_A,\,m_{H^\pm},\,\sin(\beta-\alpha),\,m_{12}^2$ and
$\tan\beta$, with $m_H$ fixed to 125\ GeV. A complete list of the
paramater ranges and the constraints imposed in the scan can be found
in\ \cite{Enberg:2016ygw}. In Fig.\ \ref{fig:ZWidth} we show the
points passing all these constraints and, additionally, lying
within the 2$\sigma$ error on the experimental
measurement of the $Z$ width. The ones highlighted in yellow are
the benchmark points (BPs) selected for a more detailed
investigation. The color map shows the production cross section for
the $q\bar{q}\to hA$ process, calculated using\ \cite{Alwall:2014hca}. 

\begin{figure}[tbp]
\begin{center}
\begin{tabular}{cc}
    \includegraphics[scale=0.38]{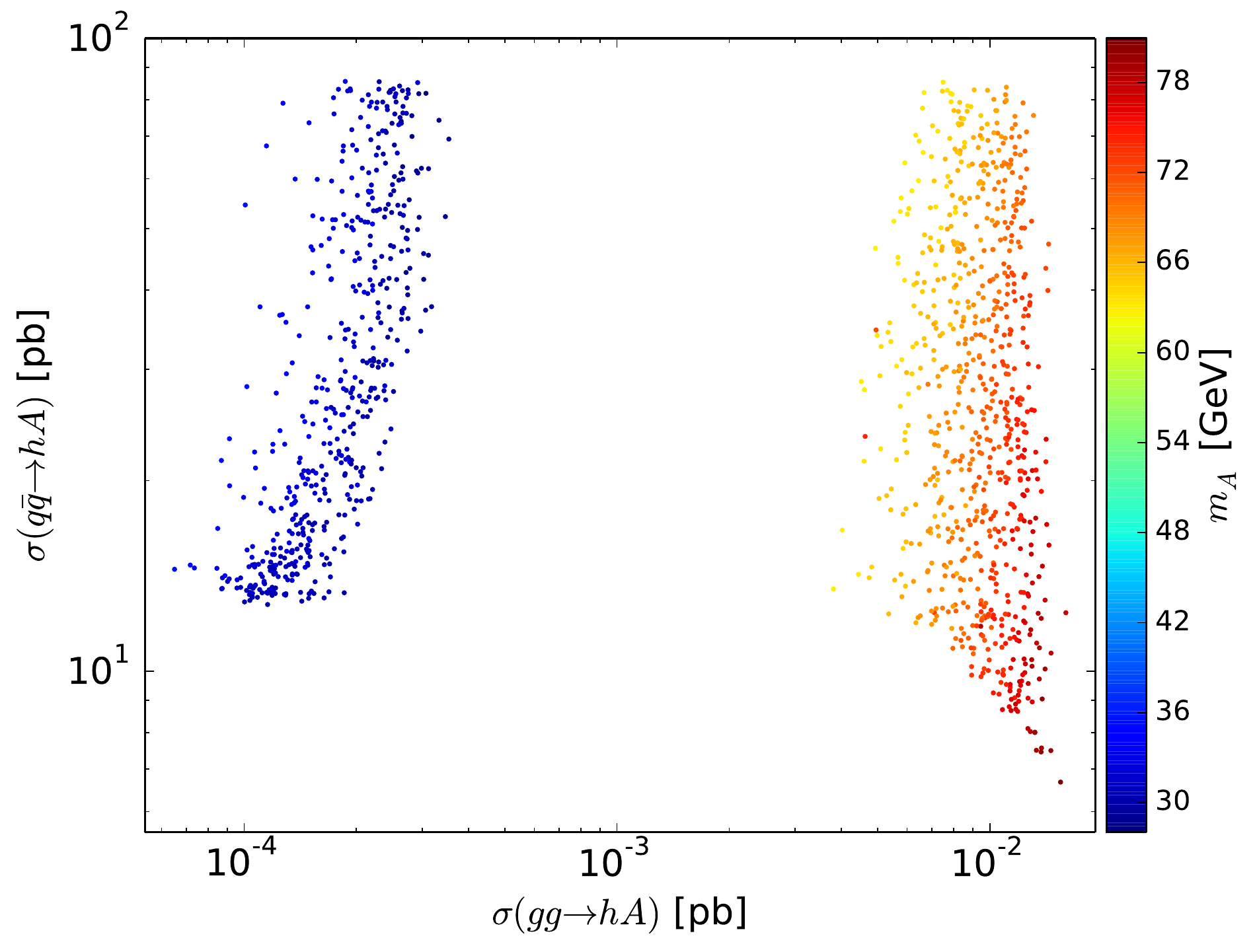}&
 \includegraphics[scale=0.38]{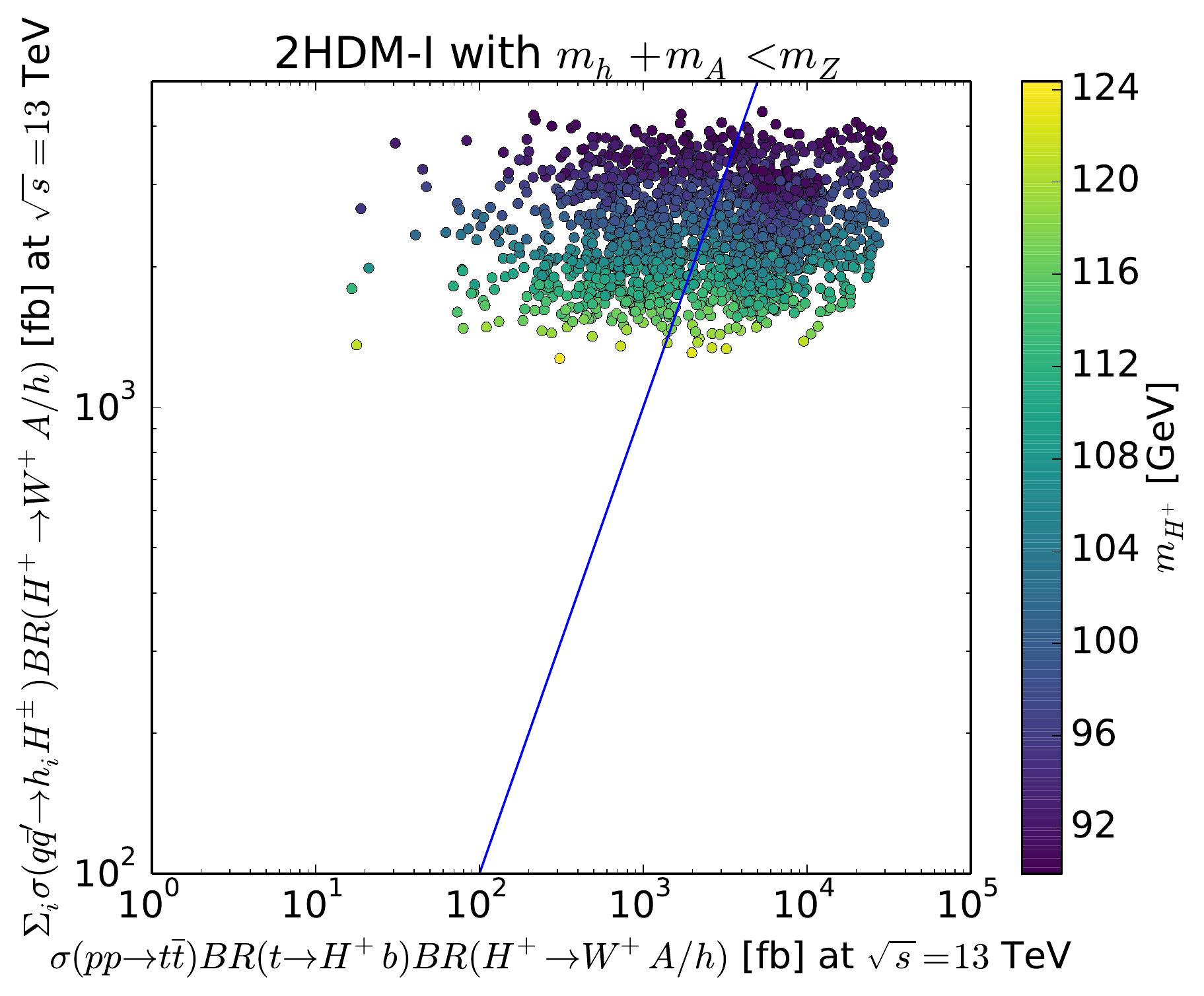}
\end{tabular}
    \caption{Left: Cross sections for QCD vs. EW production of $hA$
      pairs at the LHC, for the good points from the parameter space
      scan of Type-I 2HDM, with the color map showing the mass of $A$. 
     Right: Cross sections for two of the main
      production modes of the accompanying $H^\pm$, which decays via
      $W^\pm h$ or $W^\pm A$ and its mass is indicated by the color map.}
    \label{fig:qqvsgg}
\end{center}
\end{figure}

Fig.\ \ref{fig:qqvsgg}(left) shows that the production cross
section at the LHC with $\sqrt{s}=13$\ TeV can exceed that for the $gg\to
hA$ one, calculated using\ \cite{Hespel:2014sla}, by a few orders of magnitude, reaching up to about 90\
pb. Table\ 1 shows the cross sections corresponding to 
the two production modes for the three
BPs noted earlier. For BP1, where $m_A<m_h$, the difference between the cross
sections is much more enhanced compared to that for the other two BPs, which
correspond ot the case $m_h<m_A$.  The table also contains the
branching ratios (BRs) of the $h$ and $A$ thus produced in their two most dominant
decay modes. Clearly, when kinematically allowed, $Z^*A$ is the
primary decay channel of $h$ (for BP1) and $Z^*h$ of $A$
(for BP2 and BP3). Thus, a non-conventional final state like
$Z^*b\bar{b}b\bar{b}$ could serve as an important probe of this model scenario.    

Finally, a crucial feature of such light $h$ and $A$ is that, in order to satisfy the EW precision constraints, they are
always accompanied by a light $H^\pm$. The latter decays dominantly in the 
$W^\pm h$ or $W^\pm A$ channels, with their combined BR approaching
unity. The most significant production process(es) of $H^\pm$, which
subsequently decays in one of these two channels, can therefore have a substantial
cross section at the LHC\ \cite{hpm-2HDMI}, as shown in Fig.\
\ref{fig:qqvsgg}(right). It can thus potentially provide a
complimentary signature of the Type-I 2HDM scenario considered here. 

\begin{table}[h!]
\begin{center}
	\begin{tabular}{c c c c c c c c}%
        \hline 
		BP & $m_h$  & $m_A$ & $m_{H_\pm}$ & $\sigma (q\bar{q})$ & $\sigma (gg)$ &BR$(h\to Z^*A,\,b\bar{b})$ & BR$(A\to Z^*h,\,b\bar{b})$ \\\hline
1 & 54.2 & 33.0 & 118.3 & 41.2 & $1.5 \times
                                                        10^{-4}$ &
                                                                   0.94,\,0.05  & 0,\,0.86 \\
2 & 22.2 & 64.9 & 101.5 & 34.4 & $7.2 \times
                                                         10^{-3}$&
                                                                   0,\,0.83 & 0.86,\,0.12  \\ 
3 & 14.3 & 71.6 & 107.2 & 31.6 & $1.1 \times
                                                        10^{-2}$ & 0,\,0.60 & 0.90,\,0.08 \\
        \hline 
	\end{tabular}
    \caption{Parton-level production cross
      sections (in pb) of $h$ and $A$ pairs, and their largest
      branching ratios,
      corresponding to the three selected benchmark points.}
	\label{tab:BP}
\end{center}
\end{table}

\end{document}